\documentclass[aps,prl,twocolumn,superscriptaddress,showpacs,amsmath,amssymb,longbibliography]{revtex4-2}

\makeatletter
\AtBeginDocument{%
  \def\@bibdataout@apsrev{%
    \immediate\write\@bibdataout@bbl{\string\providecommand{\string\BIBentryALTinterwordspacing}{ }}% Handling Word Spacing
    \immediate\write\@bibdataout@bbl{\string\providecommand{\string\BIBentrySTDinterwordspacing}{ }}% Handling Word Spacing
    \immediate\write\@bibdataout@bbl{\string\providecommand{\string\BIBforeignlanguage}[2]{##2}}% Handling Foreign Languages
    \immediate\write\@bibdataout@bbl{\string\renewcommand{\string\bibnamefont}[1]{##1}}% Keep author names normal
  }%
}
\makeatother

\usepackage[T1]{fontenc}
\usepackage[utf8]{inputenc}
\usepackage{color}
\usepackage{babel}
\usepackage{amstext}
\usepackage{graphicx}
\usepackage{esint}
\usepackage{soul}
\usepackage[unicode=true,pdfusetitle,
 bookmarks=true,bookmarksnumbered=false,bookmarksopen=false,
 breaklinks=false,pdfborder={0 0 1},backref=false,colorlinks=false]
 {hyperref}
 \usepackage[normalem]{ulem}
 \usepackage{xcolor}

\makeatletter
\@ifundefined{textcolor}{}
{%
 \definecolor{BLACK}{gray}{0}
 \definecolor{WHITE}{gray}{1}
 \definecolor{RED}{rgb}{1,0,0}
 \definecolor{GREEN}{rgb}{0,1,0}
 \definecolor{BLUE}{rgb}{0,0,1}
 \definecolor{CYAN}{cmyk}{1,0,0,0}
 \definecolor{MAGENTA}{cmyk}{0,1,0,0}
 \definecolor{YELLOW}{cmyk}{0,0,1,0}
}

\usepackage{multirow}

\makeatother

\begin{document}
\title{Diffusion in interacting two-dimensional systems under a uniform magnetic field}
\author{\L ukasz Iwanek}
\affiliation{Institute of Theoretical Physics, Faculty of Fundamental Problems of Technology, Wroc\l aw University of Science and Technology, 50-370 Wroc\l aw, Poland}
\author{Marcin Mierzejewski}
\affiliation{Institute of Theoretical Physics, Faculty of Fundamental Problems of Technology, Wroc\l aw University of Science and Technology, 50-370 Wroc\l aw, Poland}
\author{Adam S. Sajna}
\affiliation{Institute of Theoretical Physics, Faculty of Fundamental Problems of Technology, Wroc\l aw University of Science and Technology, 50-370 Wroc\l aw, Poland}

\begin{abstract}
%The  dynamics of interacting particles under orbital magnetic fields are notoriously difficult to study, as their physics, in local hopping models, must be defined beyond one dimension. Here, we report on the diffusive relaxation dynamics of two-dimensional interacting fermionic systems under a uniform magnetic field in the infinite temperature regime. Using the Lanczos method, we show that the fermionic truncated Wigner approximation captures the equilibration dynamics unexpectedly well for intermediate interaction strengths when going beyond one dimension, even for relatively small ladder systems. Focusing on density-wave-like initial conditions, we discuss the dependence of the diffusion constant on different magnetic flux values per plaquette. We find that interactions of comparable magnitude that exceed the hopping energy strongly suppress magnetic-field effects on diffusive transport. However, when the interactions are comparable to the kinetic energy, diffusion is significantly reduced by flux even for relatively weak magnetic fields. This is observed for larger system sizes (above approximately 400 lattice sites), where finite-size effects weakly affect the diffusion constant. We suggest that our results should be directly accessible on current optical lattice platforms.

The dynamics of interacting particles in orbital magnetic fields are notoriously difficult to study, as this physics is inherently connected to electronic correlations in two-dimensional systems, for which no straightforward theoretical methods are available. Here, we report on the diffusive relaxation dynamics of two-dimensional interacting fermionic systems under a uniform magnetic field in the infinite temperature regime. We first show that the fermionic truncated Wigner approximation captures the equilibration dynamics unexpectedly well for intermediate interaction strengths when going beyond one dimension. This high accuracy holds at least for relatively small ladder systems, which are accessible to the Lanczos method that we use to benchmark the reliability of the Wigner approximation. We find that strong interactions, which exceed the hopping energy,  suppress magnetic-field effects on diffusive transport. However, when the interactions are comparable to the kinetic energy, the diffusion is significantly reduced by the magnetic flux. This is observed for sufficiently large systems (above approximately 400 lattice sites), where finite-size effects weakly affect particle transport. We suggest that our results should be directly accessible on current optical lattice platforms.

\end{abstract}
\maketitle

\section{Introduction}

Most simple static potentials, such as disorder or linear potentials, have challenged our understanding of equilibration in many-body lattice systems in recent years \cite{RevModPhys.91.021001, Nandkishore2015, Sierant_2025}. Thanks to readily available numerical simulations, their dynamical behavior has already been well studied in one-dimensional (1D) systems \cite{PhysRevLett.122.040606, vanNieuwenburg2019, PhysRevB.102.054206, PhysRevLett.127.240502, PhysRevB.102.104203, PhysRevResearch.2.032039, Yao2021, PhysRevB.104.014201, PhysRevB.105.L140201,Znidaric_2008, Bardarson_2012,Serbyn_2013, Andraschko2014,Mierzejewski_2016,Serbyn_2017,Sels_2021,Vidmar_2021}. Moreover, the knowledge gained in 1D is also very helpful for studies of higher-dimensional systems, where numerical tools are less efficient (see, e.g. \cite{vanNieuwenburg2019, Doggen_2020, PhysRevB.105.134204}). A number of experiments has also been performed to investigate their dynamical behavior \cite{Choi_2016,Bordia_2016,Bordia_2017,GuardadoSanchez2020, GuardadoSanchez2021, Karamlou2022, PhysRevLett.127.240502}.

Quite different situations arise in systems for which potentials can naturally be constructed only beyond one dimension. The simplest examples of such quantum systems are particles subjected to a uniform magnetic field. In these systems, additional positional degrees of freedom must be included to account for orbital effects. As a result, the relaxation dynamics of particles from a nonstationary initial state is much more difficult to study, and so far, for interacting systems, mostly the models of ladder \cite{SutharPhysRevB.101.134203, PhysRevB.106.035123, PhysRevResearch.3.013178, PRXQuantum.3.030328, mamaev2023, PhysRevA.94.043609} and next-nearest-neighbor hopping \cite{PhysRevA.94.053610} have been investigated. Current optical lattice simulators offer much greater flexibility in these regimes \cite{liviPhysRevLett.117.220401, Atala2014, Mancini2015}, including access to two-dimensional (2D) systems (see \cite{Halimeh2025} and references therein). Therefore, from the theoretical side there is a need to develop tools that can overcome the exponential growth of the Hilbert space. This has led to a situation in which the nonequilibrium dynamics of quantum many-body systems in magnetic fields remains poorly understood.

The main difficulty in studying such systems arises from the fact that, in order to properly resolve the relevant magnetic lengths, correspondingly larger system sizes must be simulated. For example, to correctly capture the dynamics of a system with, e.g., a flux of $1/6$ per plaquette, a lattice larger than $6 \times 6$ in 2D must be considered, which is already at the limit of exact numerical studies. In this work, we show that a lattice size of at least $12$ sites in one direction is required to account for finite-size effects in relaxation dynamics under a flux $\Phi = 1/6$. This naturally renders systematic studies of interacting systems out of reach for exact simulations, even for the smallest toy models. Moreover, we expect that methods relying on slow entanglement growth (see, e.g., \cite{Ors2019, Schollwck2011}) may also be inefficient because the translationally-invariant systems considered here relax quickly enough to generate a large amount of entanglement across the system. In strongly disordered systems, slow growth of entanglement enables the study of dynamics on longer timescales \cite{SutharPhysRevB.101.134203}, however, the additional effects of disorder lie beyond the scope of the present research.

In this work, we investigate the relaxation dynamics of density waves under different strengths of uniform magnetic fields. Such waves, in the context of diffusion and subdiffusion analysis, have already been studied experimentally but in the setting of tilted optical lattices \cite{GuardadoSanchez2020, Scherg2021} (which correspond to dynamics under a static electric field). Here, we focus on quasi-2D systems of interacting spinless fermions on a lattice. To study larger system sizes, we employ the fermionic truncated Wigner approximation (fTWA) \cite{Davidson_2017} and show that it provides a reasonable effective description of many-body dynamics under magnetic fields for intermediate interaction strengths. The effectiveness of fTWA is demonstrated by benchmarking it against the Lanczos method. While fTWA performs poorly in 1D, especially for integrable models, it reveals an unexpected agreement beyond one dimension. It is worth mentioning that within fTWA the magnetic fields are treated exactly, whereas the interactions are treated approximately. However, we show that interactions exceeding the hopping energy (but of comparable magnitude) are still reasonably well captured by fTWA. This enables us to analyze equilibration dynamics under various magnetic-field strengths, lattice geometries, and interaction values. For example, having established that the dynamics is diffusive, we study the diffusion coefficient and its dependence on the magnetic field. We also discuss the importance of finite-size effects, which is particularly relevant for smaller values of the flux per plaquette, where resolving the magnetic length is essential for a quantitative description of the dynamics.

It is also important to emphasize that numerical studies investigating the diffusive dynamics of interacting fermions in various magnetic fields are rather rare.  One possible explanation is that, to observe a diffusion constant in finite-size analyses at vanishing magnetic field strength, one must consider system sizes beyond the reach of exact numerical methods (as we argue in this paper). In the case of noninteracting particles, 2D lattice systems with strong magnetic fields and relatively weak disorder have already been analyzed in Ref. \cite{PhysRevLett.72.713}. The authors report that the variance of a wave packet initialized in the middle of the Landau band spreads diffusively, which is consistent with our findings.

The rest of the paper is organized as follows. In Sec.~\ref{ftwa}, we introduce the model and the fTWA method. In Sec.~\ref{initial-state-preparation} we explain initial state preparation process and how we analyze diffusion. Then, in Sec.~\ref{benchmark} we benchmark fTWA results against exact dynamics. In Sec.~\ref{diffusion in 2D}, we present the main results of our work, which concern the diffusive dynamics of interacting fermions under a uniform magnetic field. The paper concludes with Sec.~\ref{conclusions}.

\begin{figure}
\includegraphics[scale=0.92,page=1]{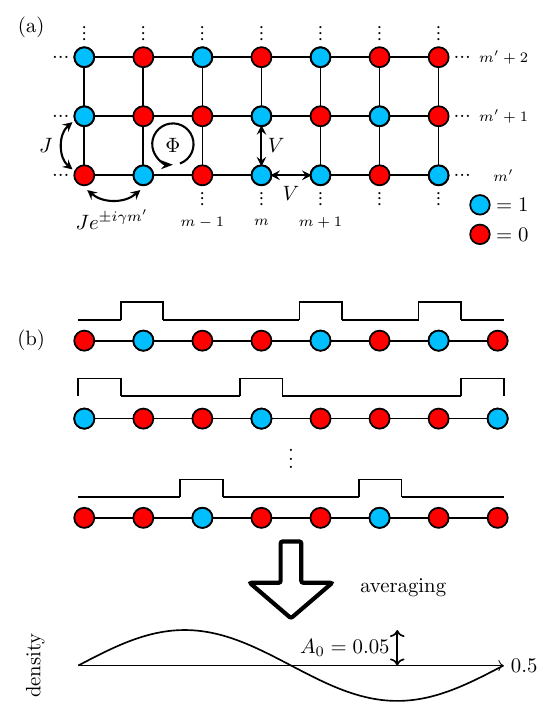}
\caption{Schematic visualization of (a) the system described in Eq.~\eqref{hamiltonian}, where blue (red) dotes show an occupied (unoccupied) sites in the 2D lattice. The magnetic flux per plaquette is given by $\Phi=\gamma/2\pi$. (b) Construction of the sine-profile density by averaging over an ensemble of initial states.}
\label{fig1}
\end{figure}

\section{Model and method} \label{ftwa}
In this paper we consider spinless interacting fermions on the lattice with the following Hamiltonian
\begin{equation}
\begin{aligned}
\label{hamiltonian}
	\hat{H}&=-\frac{J}{2}\sum_{\mathbf{r}}\left( e^{-i\gamma \mathbf{r}\cdot \mathbf{\hat{y}}}\hat c_{\mathbf{r+\hat{x}}}^{\dagger}\hat c_{\mathbf{r}} + \hat c_{\mathbf{r+\hat{y}}}^{\dagger}\hat c_{\mathbf{r}}+ h.c. \right)  \\
    &+ V\sum_{\substack{\mathbf{r} \\ \mathbf{\delta r \in {\hat{x}, \hat{y}}} }}\hat n_{\mathbf{r}}\hat n_{\mathbf{r}+\mathbf{\delta r}},
\end{aligned}
\end{equation}
where $J$ is hopping integral, $V$ is the nearest-neighbor interaction potential, $\hat c^{\dagger}_{\mathbf{r}}$ ($\hat c_{\mathbf{r}}$) is the fermionic creation (annihilation) operator at the site $\mathbf{r}$ and $\hat n_{\mathbf{r}}=\hat c^{\dagger}_{\mathbf{r}}\hat c_{\mathbf{r}}$ is an occupation operator. Throughout the paper, we analyze different values of $V/J$, taking $J = 1$ as the reference energy scale for the numerical simulations. Moreover, we define $\mathbf{r}=(m,m')$ with $m=1,\dots,L_x$ and $m'=1,\dots,L_y$ where $L_x$ and $L_y$ correspond to the lattice sizes in the $x$ and $y$ directions, respectively. The unit vectors $\mathbf{\hat{x}}$ and $\mathbf{\hat{y}}$ correspond to the shifts $(1,0)$ and $(0,1)$. In order to take into account the effects of magnetic fields, we incorporate the Peierls phase in the hopping term along the $x$-direction \cite{Goldman_2014}. The Landau gauge is used, in which the strength of the uniform magnetic field is controlled by the parameter $\gamma$. In this gauge, the flux $\Phi$ per single plaquette is given by $\gamma/2\pi$ (see Fig.~\ref{fig1}a), and the strongest magnetic field is obtained for $\gamma = \pi$.  In the following, a periodic boundary condition in the $x$-direction is used, and for the purposes of this work the system size will be denoted as $L=L_x\times L_y$. Moreover, the lattice spacing and Planck constant are set to unity throughout the manuscript.

To efficiently simulate 2D lattice systems, a fermionic version of the truncated Wigner approximation is used (so-called fTWA) \cite{Davidson_2017}. The method is based on the Wigner-Weyl formalism which allows phase space representation of quantum dynamics. Thanks to its polynomial scaling with system size, fTWA enables simulations of hundreds of lattice sites, at the cost of approximately treating interactions between particles \cite{Polkovnikov_2010}. Therefore, in the following section, we validate the fTWA by benchmarking it against exact simulations (at short times, fTWA is expected to produce asymptotically exact results by construction \cite{Polkovnikov_2010}).

The fTWA methodology for spinless fermions can be briefly described in the three main steps:  

(i) In the Wigner-Weyl formalism, fermionic degrees of freedom are represented as phase space variables $\rho_{\mathbf{r}, \mathbf{r'}}$ which are the Weyl symbol of $(\hat c^{\dagger}_{\mathbf{r}}\hat c_{\mathbf{r'}}- \hat c_{\mathbf{r'}}\hat c^{\dagger}_{\mathbf{r}})/2 \equiv \hat E^{\mathbf{r}}_{\mathbf{r'}}$ operators. In general any operator $\hat{\mathcal{O}}$ in Hilbert space, which can be written in bilinear form using $\hat E^{\mathbf{r}}_{\mathbf{r'}}$ can also be easily represented in fermionic phase space by using complex variables $\rho_{\mathbf{r},\mathbf{r'}}$, i.e. there exists a correspondence $\hat{\mathcal{O}} (\hat E^{\mathbf{r}}_{\mathbf{r'}}) \leftrightarrow \mathcal{O}_W (\rho_{\mathbf{r},\mathbf{r'}})$ (here index $W$ denotes the Weyl symbol of $\hat{\mathcal{O}}$ in phase space). As an example, occupation operator $\hat n_{\mathbf{r}}$ has Weyl symbol $\rho_{\mathbf{r},\mathbf{r}}+1/2$ and Hamiltonian's Weyl symbol [Eq.~\eqref{hamiltonian}] can be written as
\begin{equation}
\begin{aligned}
    H_{W} = &-\frac{J}{2}\sum_{\mathbf{r}} \left(e^{-i\gamma \mathbf{r}\cdot \hat{y}} \rho_{\mathbf{r},\mathbf{r}+\mathbf{\hat{x}}} +\rho_{\mathbf{r},\mathbf{r}+\mathbf{\hat{y}}}+c.c.\right)\\
    &+\frac{1}{2} V\sum_{\mathbf{r}}\left(\rho_{\mathbf{r},\mathbf{r}}+\frac{1}{2}\right)^2 
    \\&+ V\sum_{\substack{\mathbf{r} \\ \mathbf{\delta r \in {\hat{x}, \hat{y}}} }}\left(\rho_{\mathbf{r},\mathbf{r}}+\frac{1}{2}\right)\left(\rho_{\mathbf{r}+\mathbf{\delta r}, \mathbf{r}+\mathbf{\delta r}}+\frac{1}{2}\right), 
    \label{eq: H_W}
\end{aligned}
\end{equation}
see also Ref. \cite{Iwanek_2023} for comparison where magnetic-free Hamiltonian was considered.

(ii) Quantum evolution in phase space is governed by Hamilton's equations of motion

\begin{equation} \label{Hamilton eqns}
	i\frac{d\rho_{\mathbf{r}, \mathbf{r'}}}{d t} = \sum_{\mathbf{r''}} \left({\partial H_W\over \partial \rho_{\mathbf{r'},\mathbf{r''}} } \rho_{\mathbf{r},\mathbf{r''}}-{\partial H_W\over \partial \rho_{\mathbf{r''},\mathbf{r}}} \rho_{\mathbf{r''},\mathbf{r'}}\right),
\end{equation}
where the initial conditions $\rho_{\mathbf{r}, \mathbf{r'}}(t=0) \equiv \rho_{\mathbf{r}, \mathbf{r'}}^{0}$ are sampled from the initial Wigner function $\mathcal{W}(\rho^{0})$. For example, when starting the quantum evolution from a product state, the Wigner function can be approximated as a product of Gaussians \cite{Davidson_2017}
\begin{equation}
\mathcal{W}(\rho^{0})=\prod_{\mathbf{r}, \mathbf{r'}} \frac{e^{-(\rho_{\mathbf{r}, \mathbf{r'}}^{0}-\mu_{\mathbf{r}, \mathbf{r'}})((\rho_{\mathbf{r}, \mathbf{r'}}^{0})^{*}-\mu_{\mathbf{r}, \mathbf{r'}})/2\sigma_{\mathbf{r}, \mathbf{r'}}^{2}}}{\sqrt{2\pi \sigma_{\mathbf{r}, \mathbf{r'}}^{2}}},
\end{equation}
where $\mu_{\mathbf{r}, \mathbf{r'}}$ and $\sigma_{\mathbf{r}, \mathbf{r'}}$ are determined by the expectation and variance of the bilinear operators $E^{\mathbf{r}}_{\mathbf{r'}}$ calculated with respect to the initial quantum state \cite{Davidson_2017}, i.e.
\begin{align}
\langle \hat{E}_{\mathbf{r}_2}^{\mathbf{r}_1}\rangle_{t=0}=\int \rho_{\mathbf{r}_1, \mathbf{r}_2}^{0}\mathcal{W}(\rho^{0}){\mathrm D}\rho^{0},
\end{align}
\begin{align}
&\frac{1}{2} \langle \hat{E}_{\mathbf{r}_2}^{\mathbf{r}_1}\hat{E}_{\mathbf{r}_4}^{\mathbf{r}_3} + \hat{E}_{\mathbf{r}_4}^{\mathbf{r}_3}\hat{E}_{\mathbf{r}_2}^{\mathbf{r}_1} \rangle_{t=0}  =\int \rho_{\mathbf{r}_1, \mathbf{r}_2}^{0}\rho_{\mathbf{r}_3,\mathbf{r}_4}^{0}\mathcal{W}(\rho^{0}){\mathrm D}\rho^{0},
\end{align}
where ${\mathrm D}\rho^{0}=\prod_{\mathbf{r}, \mathbf{r'}}{\mathrm d}\rho_{\mathbf{r}, \mathbf{r'}}^{0}$.

(iii) Expectation value of observable $\mathcal{O}$ in fTWA is given by
\begin{equation}
\left\langle \hat{\mathcal{O}}\right\rangle_{t} \approx\int \mathcal{O}_{W}[\rho(t)] \mathcal{W}(\rho^0){\mathrm D}\rho^0,\label{eq: average fTWA}
\end{equation}
which is formally obtained by averaging $\mathcal{O}_{W}[\rho(t)]$ over all phase-space trajectories at time $t$, sampled from the Wigner function $\mathcal{W}(\rho^{0})$.

\iffalse
\begin{equation}
\begin{aligned}
i\frac{d\rho_{\mathbf{r}, \mathbf{r'}}}{dt} =& -\sum_{\mathbf{r''}} \left(J_{\mathbf{r'},\mathbf{r''}}\rho_{\mathbf{r},\mathbf{r''}}-J_{\mathbf{r''},\mathbf{r}}\rho_{\mathbf{r''},\mathbf{r'}} \right) \\
&+\rho_{\mathbf{r},\mathbf{r'}}\sum_{\mathbf{r''}}\rho_{\mathbf{r''},\mathbf{r''}}\left(V_{\mathbf{r'},\mathbf{r''}}-V_{\mathbf{r},\mathbf{r''}} \right) \\
&+V\rho_{\mathbf{r},\mathbf{r'}}\left(\rho_{\mathbf{r'},\mathbf{r'}}-\rho_{\mathbf{r},\mathbf{r}} \right),
\end{aligned}
\end{equation}
\fi

\section{Initial state preparation and diffusion analysis}
\label{initial-state-preparation}
To investigate relaxation processes in the analyzed fermionic system, we study the decay of an initially prepared density wave. A similar approach was recently implemented in two-dimensional systems in experiments with cold atoms in optical lattices \cite{GuardadoSanchez2020,GuardadoSanchez2021}. 

We are interested in the diffusive dynamics which is described by the differential equation
\begin{equation}
	\frac{\partial p(x,t)}{\partial t} = D \frac{\partial^2 p(x,t)}{\partial x^2},
	\label{eq:diffusion_equation}
\end{equation}
where $D$ is diffusion constant. For periodic boundary conditions, the solution of the above equation, that describes charge excitations of a half-filled system,  can be chosen as
\begin{equation}
	p(x,t)  = 0.5 + A(t) \sin(2\pi x/\lambda) \, ,
    \label{p(x,t)}
\end{equation}
where $\lambda$ is wavelength in $x$-direction and $A(t)$ satisfy
\begin{equation}
	A(t) = A_0 \exp(-\alpha t)\, ,
	\label{eq:exp_decay_alpha}
\end{equation}
with the decay rate
\begin{equation}
	\alpha = \frac{4\pi^{2} D}{\lambda^{2}} \, .
    \label{alpha}
\end{equation}

To investigate whether the dynamics of interacting fermions follow diffusive behavior, we prepare an initial states in the form reflecting Eq.~\eqref{p(x,t)} and analyze their relaxation. Following this idea, the initial state is prepared in the form of sinusoidal density-wave with the modulation along $x$-direction. Specifically, simulations start from product states in which every site $\mathbf{r}=(m,m')$ is randomly assigned as occupied or empty and the probability for drawing an occupied state is $0.5 + A_0 \sin(2\pi m/\lambda) \in [0,1]$. We denote such a choice of initial state by $|\psi_s\rangle$ with a unique index $s$. Taking the average over a correspondingly large number of such initial states the approximate initial profile has a form (see also Fig.~\ref{fig1}b)
\begin{equation}
	\overline{\langle \hat n_{\mathbf r}(0) \rangle} = \frac{1}{N}\sum_{s=1}^{N}\langle \psi_s | \hat n_{\mathbf r}(0) |\psi_s \rangle \approx 0.5 + A_0 \sin(2\pi m/\lambda) \,,
    \label{profile}
\end{equation}
where $N$ is a number of random initial states in the ensemble. Throughout the work, we assume amplitude of density waves as $A_0=0.05$ and unless stated otherwise, we set the wavelength to the system size in the modulation direction $\lambda=L_{x}$. Moreover, for additional smoothing of density profile an averaging along density stripes (i.e. $y$-direction) is performed for each time, i.e. $(1/L_{y})\sum_{m'}\overline{\langle \hat{n}_{\mathbf (m,m')}(t)} \rangle \equiv n_m(t)$. The size of ensemble, $N$,  depends on the system geometry. Increasing $L_{y}$ reduces fluctuations because the stripe averaging involves more sites, $m'$, for each fixed $m$. This improves self-averaging of $n_{m}(t)$. In contrast, for larger $L_{x}$ and fixed $L_{y}$, a larger $N$ is needed, since the profile contains more bins in $x$-direction. To accurately reproduce the  spatial profile from Eq.~\eqref{profile} by binary random initial occupations, one needs  sufficiently large $N\sim {\cal O}(10^2)\div {\cal O}(10^3)$.

In order to check that the initial density-wave profile spreads diffusively, we use the following fitting procedure. We fit $p(x=m',t)$ from Eq.~\eqref{p(x,t)} to $n_m(t)$ to determine the dependence of $A(t)$. If an exponential decay of $A(t)$ is observed in our simulations with the decay rate \mbox{$\alpha \propto 1/L_x^2$}, we then obtain the diffusion constant $D$ using the relations from Eqs.~\eqref{eq:exp_decay_alpha}-\ref{alpha}.

\begin{figure}
\includegraphics[scale=1.033,page=1]{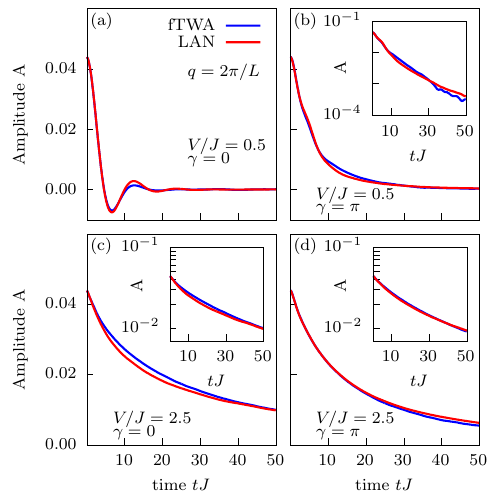}
\caption{Time evolution of the amplitude $A$ for a $10\times 2$ ladder system with $1000$ initial states, comparing two methods: Lanczos (red) and fTWA (blue). Panels (a), (c) show results without a magnetic field, while (b), (d) correspond to $\gamma=\pi$. Panels (a), (b) are plotted for interaction strength $V/J=0.5$, and panels (c), (d) for $V/J=2.5$. All panels represent data for the longest wavelength mode, i.e., $\lambda = L_x= 10$ in units of the lattice spacing in the $x$-direction. For fTWA, $5000$ trajectories were sampled.}
\label{fig2}
\end{figure}

\begin{figure}
\includegraphics[scale=1.03,page=1]{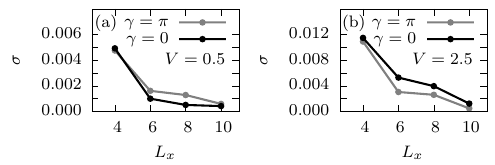}
\includegraphics[scale=1.03,page=1]{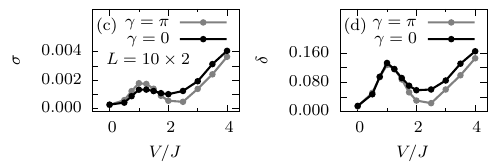}
\caption{The corresponding deviations of fTWA from the Lanczos results are shown for different system sizes $L_x$ (a)-(b) and interaction strengths (c)-(d). Here, $\sigma$ and $\delta$ denote the absolute and relative errors, respectively (see Eqs.~\eqref{absolute}-\eqref{relative}). The system and simulation parameters correspond to those used in Fig.~\ref{fig2}, and the errors are calculated for the time range $(0,T)=(0,50)$ in units of $J$.}
\label{fig3}
\end{figure}

\begin{figure}[t]
\includegraphics[scale=1.06,page=1]{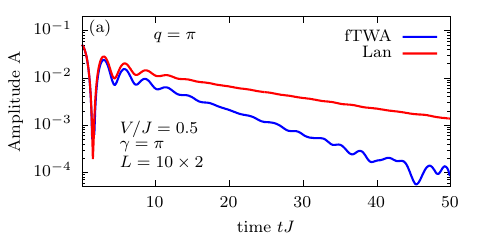}
\includegraphics[scale=1.06,page=1]{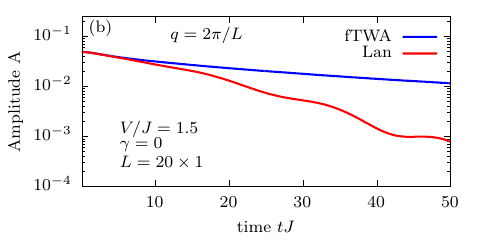}
\caption{Time evolution of the amplitude $A$ for (a) $10\times 2$ ladder system and (b) $20\times 1$ one-dimensional system, comparing Lanczos (red) and fTWA (blue). Panel (a) shows the shortest wavelength mode $q=\pi$ with interaction strength $V/J=0.5$ and $600$ initial states, while panel (b) displays the longest wavelength mode $q=2\pi/L$ with interaction strength $V/J=1.5$ and $5000$ initial states. For fTWA, $5000$ trajectories were sampled.}
\label{fig4}
\end{figure}

\begin{figure*}[t]
\includegraphics[scale=1.06,page=1]{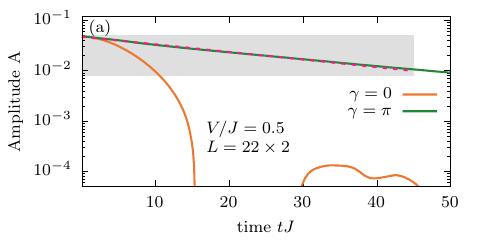}
\includegraphics[scale=1.06,page=1]{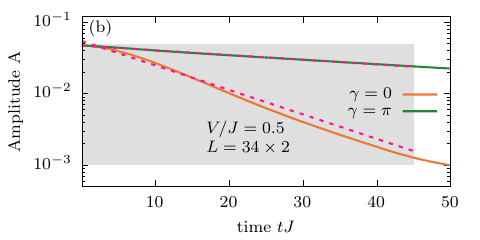}
\includegraphics[scale=1.06,page=1]{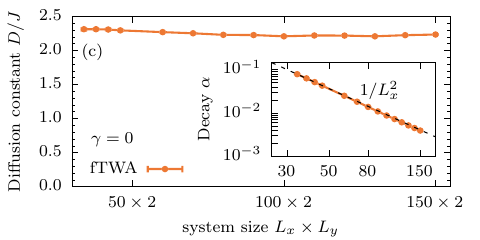}
\includegraphics[scale=1.06,page=1]{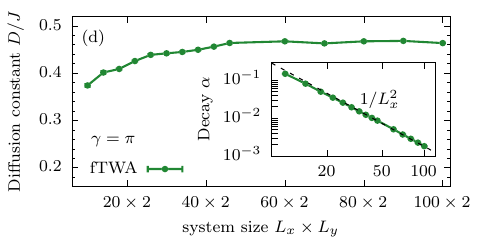}
\caption{Panels (a) and (b) show the time evolution of the amplitude $A$ for system size $L=22\times 2$ and $L=34\times 2$, respectively, with phase $\gamma=0$ and $\gamma=\pi$. Panels (c) and (d) present the diffusion constant $D/J$ as a function of system size $L_x$: panel (c) without a magnetic field, $\gamma=0$ and panel (d) with $\gamma=\pi$. In panels (a) and (b), the red dashed line indicates the fitted function $\exp(-\alpha t)$, with the fitting region marked by a gray rectangle spanning $t \in [0;45]$. The insets in panels (c) and (d) display the coefficient $\alpha$ as a function of system size $L_x$, obtained from fitting $\exp(-\alpha t)$ in the same time window $t \in [0;45]$. The number of initial states is chosen as $100 \times L_{x}$ (e.g., $34\times 2$ corresponds to $3400$ states). For fTWA, the number of trajectories varies from $5000$ for the smallest system to $100$ for the largest system.}
\label{fig5}
\end{figure*}

\begin{figure}
\includegraphics[scale=1.06,page=1]{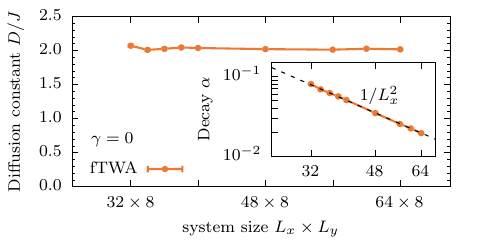}
\caption{The diffusion constant $D/J$ as a function of system size $L$ for system sizes between $32 \times 8$ and $64 \times 8$. The remaining parameters are the same as in Figs.~\ref{fig5}(c) and (d). }
\label{fig6}
\end{figure}

\begin{figure*}
\includegraphics[scale=1.06,page=1]{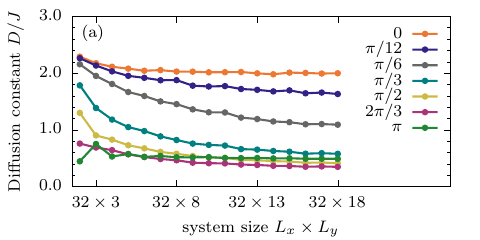}
\includegraphics[scale=1.06,page=1]{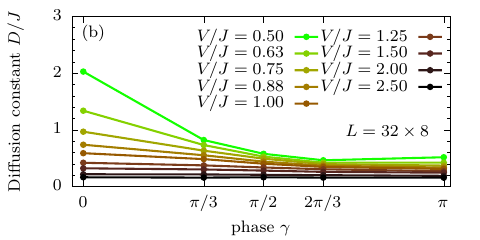}
\includegraphics[scale=1.06,page=1]{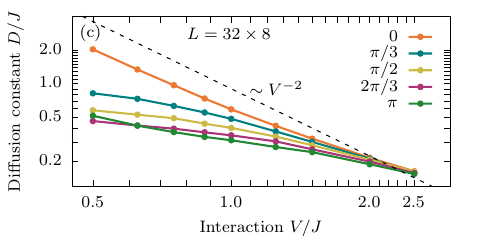}
\caption{Diffusion constant $D/J$ as a function of (a) system size $L$, (b) magnetic phase $\gamma$ and (c) interaction strength $V/J$. In panel (a), $L$ ranges from $32\times 2$ to $32 \times 18$, with curves corresponding to different values of $\gamma$. Panel (b) shows $D/J$ as a function of $\gamma$ for interaction strengths $V/J \in [0.5;2.5]$. Panel (c) shows $D/J$ on a logarithmic scale as a function of $V/J$ for selected values of $\gamma$. The number of initial states $N$ and fTWA trajectories $T$ decrease with increasing $L_{y}$ size, ranging from $N=5000$ and $T=1000$ for ($32\times 2$) and $N=260$ and $T=100$ for ($32\times 18$).}
\label{fig7} 
\end{figure*}

\section{fTWA benchmark}\label{benchmark}

To benchmark the fTWA method against exact simulations, we use the Lanczos method to compute the accurate dynamics \cite{Park_1986, Mierzejewski_2010} and study a ladder lattice of size $10 \times 2$. The decay of the amplitude is shown in Fig.~\ref{fig2}, demonstrating that fTWA provides a reasonable description of many-body dynamics in magnetic fields during relaxation processes.  The results without a magnetic field are plotted in Figs.~\ref{fig2}(a) and (c) for the interactions strength $V/J=0.5$ and $V/J=2.5$, respectively, while the results with a magnetic field $\gamma=\pi$ are presented in Figs.~\ref{fig2}(b) and (d) with the corresponding interaction strength $V/J=0.5$ and $V/J=2.5$, respectively. Moreover, it is important to note that error produced by fTWA systematically drops for larger system sizes, see the deviation $\sigma$ in Figs.~\ref{fig3}(a) and (b). The deviation between both methods, $\sigma$, is quantified as
\begin{equation}
\begin{aligned}
    \sigma = \sqrt{\frac{1}{T}\int_0^T  \left(A^{\text{Lan}}(t)-A^\text{fTWA}(t)\right)^{2}dt},
\label{absolute}
\end{aligned}
\end{equation}
where $A^\text{fTWA}(t)$ [$A^\text{Lan}(t)$] corresponds to amplitudes calculated using the fTWA (Lanczos) method.  The behavior of $\sigma$, such as in Fig. \ref{fig3},  is common in semiclassical descriptions, as an increased number of spatial or internal degrees of freedom can lead to the suppression of higher order quantum corrections in the dynamics \cite{Polkovnikov_2010,Osterkorn2024, Iwanek2025}. We also verify that, for interaction strengths up to $V/J\approx2.5$, the fTWA method reproduces the exact dynamics reasonably well (see Fig.~\ref{fig3}(c)). Moreover, due to the vanishing amplitude $A(t)$ at longer times, we also calculate the relative error for comparison, which reads
\begin{equation}
\begin{aligned}
    \delta = \sqrt{\frac{\int_0^T  \left(A^{\text{Lan}}(t)-A^\text{fTWA}(t)\right)^{2}dt}{\int_0^T (A^{\text{Lan}}(t))^{2}dt}}.
\label{relative}
\end{aligned}
\end{equation}
The results for $\delta$ are plotted in Fig.~\ref{fig3}(d), which exhibit behavior similar to the standard deviation presented in Fig.~\ref{fig3}(c). This is consistent with the fact that interactions in the truncated Wigner method are treated only approximately and are inadequate for studying systems at larger interaction strengths \cite{Polkovnikov_2010}.

It is worth noting that in the magnetic-field regime $\gamma = \pi$, the Lanczos simulation already shows that the amplitude relaxes nearly exponentially, see insets in Figs.~\ref{fig2}(b)-(d). This indicates that diffusive behavior can be resolved even in a relatively small $10 \times 2$ system for strong magnetic fields. However, we also show that finite-size effects remain significant in  smaller systems. We examine this behavior in greater detail for larger systems in the next section.

Moreover, our simulations are performed in the regime where density waves possess a weak amplitude and a long wavelength relative to the lattice size. This limit ensures reliable fTWA results over longer time scales. In contrast, for short wavelengths (see Fig.~\ref{fig4}(a)), TWA decays faster than the Lanczos prediction, as has been observed in a number of studies of both disordered and disorder-free systems in which shorter wavelengths are probed by imbalance functions \cite{PhysRevA.96.033604, Wurtz2018,Iwanek_2023, Kaczmarek_2023}. The dimensionality of the lattice plays a crucial role in fTWA predictions, except  the case of very weak interaction. Specifically, in one dimension (no magnetic field) and for $V/J\gtrsim0.5$, we observe a pronounced deviation of fTWA from the exact dynamics, suggesting that the strongly interacting regime of a one-dimensional system lies beyond the reach of the semiclassical method analyzed here. The reason for this may be that the system represented by the 1D interacting spinless Hamiltonian with local hopping and interactions is integrable and the fTWA does not properly describe the corresponding integrals of motion. From this point onward, we present results only for the longest wavelength mode, $q=2\pi/L_x$, that is relevant for the diffusive transport. 

\section{Diffusion in two-dimensional interacting systems with magnetic field \label{diffusion in 2D}}

In this section, we carefully analyze the influence of the magnetic field on the system relaxation dynamics. Different system sizes, magnetic fields, and interaction strengths are considered within the regime where fTWA produces reliable results at long times (see Sec.~\ref{benchmark}).

For weaker interactions, the equilibration of the density wave is faster, therefore, we begin our analysis by examining the impact of finite-size effects on the quantum dynamics. We focus on ladder systems and set the interaction strength to $V/J=0.5$. For relatively small system sizes ($L=22\times2$), the dynamics under a magnetic field ($\gamma=\pi$) exhibit a much slower decay of amplitude compared to the zero-field case ($\gamma=0$), see Fig.~\ref{fig5}(a). For the chosen magnetic field, the amplitude decays already exponentially, whereas in the absence of a field we observe strong finite-size dependence (cf. Figs.~\ref{fig5}(a) and (b)). Our numerical data suggest that the zero-field relaxation acquires an exponential character for moderate lattice sizes $L_x \gtrsim 34$ (Fig.~\ref{fig5}(b)). Such exponential decay is expected in diffusive systems, for which the amplitude $A(t)$ of the density wave is predicted to decay as $\sim\exp(-4\pi^2 D t / L_x^2)$, see Sec.~\ref{initial-state-preparation} for a detailed discussion. The diffusion constant $D$ extracted for different ladder sizes is plotted in Figs.~\ref{fig5}(c) and (d), both in the absence ($\gamma=0$) and in the presence ($\gamma=\pi$) of a magnetic field. Representative fits to the exponential behavior are shown as dashed lines in Fig.~\ref{fig5}(a) and (b). It is clearly seen in Figs.~\ref{fig5}(c) and (d) that presented curves saturate for analyzed system sizes enabling extraction of diffusion coefficients for ladder geometry with $D/J\approx 2$ without magnetic field and $D/J\approx 0.5$ with $\gamma=\pi$. These results suggest that for $\gamma=\pi$ diffusion constant is significantly smaller than in no-field regime showing that magnetic fields strongly suppress the diffusive relaxation.

We verify that the decay rate of the density-wave amplitude for systems beyond the ladder geometry also scales as $\sim 1/L_x^2$. As an example, we plot quasi-2D system sizes in the no-field limit (see Fig.~\ref{fig6}). 

Having established the impact of finite-size effects in the two limiting cases of magnetic-field strength, i.e., $\gamma = 0$ and $\gamma = \pi$, we now focus our attention purely on the diffusion constant and its dependence on various system sizes and field strengths. In Fig.~\ref{fig7}(a) we show finite-size effects for intermediate values of $\gamma$. The data  clearly indicate a pronounced suppression of the diffusion constant $D$ when transitioning from the ladder lattice to the 2D system for $\gamma \in (0,\pi)$. Such a drift of the diffusion constant for intermediate magnetic field values can be explained by the fact that larger systems are required to fully account for fractional fluxes per plaquette in all directions of the 2D lattice. Our data suggest that for $\gamma\gtrsim\pi/3$ the value of $D$ stabilizes above the $L=32 \times 12$ lattice size, and that the diffusion constant is reduced by at least a factor of four when the analyzed magnetic field strengths are applied. For smaller field strengts ($0<\gamma\lesssim\pi/3)$, the diffusion constant is still relaxing, indicating significant finite size effects. Therefore, the obtained data show the important role of system sizes, which have to be considered to fully resolve magnetic field effects. For example, diffusion constant starts to saturate for sizes above 12 in $y$ direction when flux $\Phi=1/6$ (i.e. $\gamma=\pi/3$) is considered (the size in the y direction is twice as large as the maximum distance a particle needs to hop in this direction in order to acquire the full flux per plaquette). %(the size in $y$ direction is two times larger than particle needs to hop in this direction in order to gain the whole flux per plaquette). 

Finally, we focus on the role of interaction strength in our diffusive dynamics. According to the fTWA benchmark given in Sec.~\ref{benchmark}, we restrict the interaction strength to $V/J \lesssim 2.5$. From the data shown  in Figs.~\ref{fig7}(b)-(c), we observe that increasing interaction strongly suppresses diffusion in our 2D system, with a more pronounced effect in the low magnetic field regime. The dashed line in Fig.~\ref{fig7}(c) shows a dependence, \mbox{$D \propto V^{-2}$}, which one expects from the assumption that the scattering rate grows as  $V^{2}$. For stronger interactions ($V/J > 1$), the magnetic field has a weak effect on the diffusion constant, suggesting that interactions play a significant role in the diffusive dynamics.

\section{Conclusions} \label{conclusions}
In this work, we study relaxation dynamics of 2D interacting spinless fermions under uniform magnetic fields. We show that semiclassical method known as a fTWA, unexpectedly well capture the diffusive quantum many-body dynamics at the system sizes which enable systematic studies of the magnetic field effects. Its effectiveness is demonstrated by benchmarking it against the Lanczos method, showing surprisingly good agreement beyond one dimension, even for small interacting ladder systems.

Motivated by recent experiments in ultracold atoms on optical lattices \cite{GuardadoSanchez2020, GuardadoSanchez2021, Scherg2021} we investigate relaxation dynamics of density waves in Landau gauge. Systematic studies of observed diffusive dynamics are performed. We demonstrate that presence of magnetic fields clearly distinguish two types of transport in the systems, i.e. with and without magnetic fields. Namely, for studied magnetic field strengths the relaxation dynamics is strongly suppressed, which shows up in comparable values of diffusion constant when field is turned on. However, to observe this effect, correspondingly large system sizes are needed to efficiently resolve magnetic length. 

Moreover, thanks to fTWA abilities to simulate intermediate interaction strength, we also investigate the influence of interaction on the quantum dynamics. We show that magnetic field effects on diffusive dynamics are relatively strong for interactions that are weaker than or comparable to the hopping energy. However, interactions above this value cause the role of the magnetic-field length scale to be strongly suppressed, and the orbital effects arising from the attached flux can be neglected in our diffusive analysis.

We suggest that currently available optical lattice systems with ultra cold atoms are able to verify our findings. Both uniform magnetic fields \cite{aidelsburgerPhysRevLett.107.255301, Halimeh2025} and density-waves \cite{GuardadoSanchez2020, GuardadoSanchez2021, Scherg2021} have been already studied in optical lattice systems.

\section{Acknowledgments} \label{acknowledgments}
We thank Anatoli Polkovnikov for comments and discussions. Numerical studies in this work have been carried out using resources provided by the Wroclaw Centre for Networking and Supercomputing \cite{wcss}, Grant No. 551 (1692966935) and Grant No. 1721138190.

%\bibli ographystyle{apsrev4-2}
\bibliography{library.bib}

\end{document}